\documentclass{webofc}
\usepackage[varg]{txfonts}   
\begin{document}
\title{Cosmological Lithium Problems}
%
%

\author{\firstname{C.A.} \lastname{Bertulani}\inst{1}\fnsep\thanks{\email{carlos.bertulani@tamuc.edu}} \and
        \firstname{Shubhchintak} \lastname{}\inst{1}\fnsep\thanks{\email{khajuria1986@gmail.com}} \and
        \firstname{A.M.} \lastname{Mukhamedzhanov}\inst{2}\fnsep\thanks{\email{akram@comp.tamu.edu}}
}

\institute{Department of Physics and Astronomy, Texas A\&M University-Commerce, Commerce, TX 75429, USA
\and
          Department of Physics and Astronomy, Texas A\&M University, College Station, TX 77843, USA 
          }

\abstract{%
We briefly describe the cosmological lithium problems followed by a summary of our recent theoretical work on the magnitude of  the effects of electron screening, the possible existence of  dark matter parallel universes and  the use of non-extensive (Tsallis) statistics during big bang nucleosynthesis. Solutions within nuclear physics are also discussed and recent measurements of cross-sections based on indirect experimental techniques are summarized. 
}
\maketitle
\section{Introduction}
\label{intro}

The cosmological lithium problem has become one of the most intriguing open questions in cosmology due inconsistencies between observation and calculations based on the standard Big Bang nucleosynthesis (BBN) for the primordial elemental abundances. The BBN model contains a few parameters such as the baryon-to-photon ratio $\eta = n_b/n_\gamma$, the neutron decay time $\tau_n$, and the number of neutrino families $N_\nu$ (see, for instance, Ref. \cite{BK16}).  The parameter $\eta$ relates to the baryon density of the universe by means of $\Omega_0 h^2  \simeq (\eta/10^{-10})/273$, with the Hubble dimensionless parameter $h$ defined through the relation $H_0=100 h$ km/s/Mpc, the index `0' meaning present time.  The anisotropies of the cosmic microwave radiation (CMB) independently determine the value of $\eta$ \cite{,Kom11,Ade16} when the universe was about  0.3 Myr after the Big Bang. Then photons decoupled and began steaming freely in the universe. Precise LEP experiments to deduce the number of neutrino families  \cite{ALE06} lead to the value $N_\nu = 2.9840 \pm 0.0082$, and neutron lifetime measurements have inferred that $\tau_n \simeq 880.2 \pm 1.0$ s \cite{Pat16}.

The observed abundances of light elements probe the universe at the very early stages, i.e., 3-20 minutes, of its existence. During this epoch,  the light elements D, $^3$He, $^4$He, and $^7$Li were produced and their abundances in selected astrophysical environments are telltales of the BBN epoch. The BBN model predictions also depend on the nuclear reaction network and magnitude of the nuclear cross sections. A few minutes ($\sim 3 $ min) after the Big Bang,  deuterons were formed  by neutron capture on protons, by means of the reaction p(n,$\gamma$)d. The formation of deuterons is strongly dependent on the  value of $\eta$.  Deuterons are promptly destroyed once they are formed leading to the formation of $^3$He nuclei by means of the (p,$\gamma)^3$He and d(d,n)$^3$He reactions. Deuterons also synthesize tritium  by means of the d(d,p)t reaction. $^4$He are then created by the $^3$He(d,p)$^4$He and t(d,n)$^4$He reactions. In the end, the BBN model predicts that the universe should be composed of about 75\% of hydrogen and 25\% of helium with tiny traces of D, $^3$He, $^7$Li and $^6$Li.   The foundations of these results rely on the big bang prediction of the neutron-to-proton ratio n/p = 1/7 when the nucleosynthesis started, i.e., the BBN occurred in a proton-rich environment. 

\begin{center}
\begin{table}[htbp]
\vspace{0.0cm}
\begin{center}
\begin{tabular}{|l|l|l|l|l|}
\hline
\hline
${\rm n} \leftrightarrow {\rm p}$  &${\rm p(n,}\gamma{\rm )d}$  & ${\rm d(p,}\gamma)^3{\rm He}$ & ${\rm d(d, p)t}	$ \\ \hline\hline
${\rm  d(d, n)}^3{\rm He}	$ &$ ^3{\rm He(n,p)t}$  &${\rm  t(d,n)}^4{\rm He}$ & $^3{\rm He(d, p)}^4{\rm He}  $  \\ \hline
$^3{\rm He}(\alpha,\gamma)^7{\rm Be}$  & ${\rm t(}\alpha,\gamma)^7{\rm Li}$ & $^7{\rm Be(n,p)}^7{\rm Li}$  &$^7{\rm Li(p,} \alpha )^4{\rm He}	$ \\ \hline
\hline
\end{tabular} 
\end{center} 
\caption{Nuclear reactions of importance for big bang nucleosynthesis. }\label{rent} 
\end{table} 
\end{center} 

The standard BBN model predicts  the $^7$Li/H abundance ratio of the order of $10^{-10}$ and the $^6$Li/H abundance ratio of the order of $10^{-14}$.  Much after the BBN epoch,  $^6$Li can  be produced in spallation processes by cosmic rays and $^7$Li can be synthesized in novae or  during AGB stars pulsations.  In Ref. \cite{Spi82} it was reported that the $^7$Li abundance is independent of the metallicity in metal-poor stars with small Fe/H abundances relative to the sun. Such stars are warm ($5700 \leq T  \leq 6250$ K)  metal-poor dwarf stars observed in the galaxy halo. For low metallicity stars the $^7$Li abundance is nearly constant and this behavior is known as the ``Spite plateau'' \cite{Spi82}. $^7$Li is destroyed in red giants with core temperatures in excess of  $10^6$ K via the reaction ${}^{7}{\rm Li}({\rm p},\alpha){}^{4}{\rm He}$ and that is why white dwarfs at moderate temperatures have been used in such observations.  The Spite plateau provides a reasonable evidence that lithium is neither created nor destroyed in warm dwarfs and that such stars display the abundances of  primordial $^7$Li.  On the other hand, it is worthwhile mention that recent observations in low-metallicity stars seem to contradict the conclusions drawn from the Spite plateau \cite{Aok09,Sbo10}. The currently accepted value for the $^7$Li BBN model abundance, calculated using $\eta = (6.07 \pm 0.07) \times 10^{-10}$
\cite{Ade16}, corresponds to  ${\rm Li/H} = (4.16 - 5.34) \times 10^{-10}$ \cite{Piz14} while the observations from metal-poor halo stars yields ${\rm Li/H}  = (1.58+ 0.35-0.28) \times 10^{-10}$ \cite{Sbo10,Coc12}.  This is approximately a factor 3 lower than expected and is the source of the lithium puzzle.

 The second lithium puzzle involves the abundance of ${}^{6}{\rm Li}$  produced during the BBN by means of the $^{2}{\rm H}(\alpha,\gamma)^{6}{\rm Li}$ reaction. $^{6}{\rm Li}$ nuclei formed in stars disappear quickly by means of other reactions. $^6$Li is also created  in  cosmic ray interactions, and could also exist in the atmosphere of  metal-poor warm dwarfs in the halo of the galaxy, surviving destruction by cosmic rays. But such assumptions are controversial, because they can  also apply to ${}^{7}{\rm Li}$ nuclei. The second lithium puzzle relates to the BBN predictions of the isotopic ratio  ${}^{6}{\rm Li}/{}^{7}{\rm Li}  \sim  10^{-5}$ \cite{Piz14,MSB16},  while observations report $^{6}{\rm Li}/{}^{7}{\rm Li}  \sim 5 \times 10^{-2}$ \cite{Asp06}.  This puzzle is less robust because of the complexities involved in 3-dimension calculations involving convection and non-local thermodynamical equilibrium, in particular in the photo-sphere of metal-poor stars. Because they might have a large influence on the $^{6}{\rm Li}/{}^{7}{\rm Li}$ isotopic ratio,  such complexities weakens the arguments for the existence of the second lithium problem which in some scenarios yields a better agreement with BBN predictions \cite{LMA13}.  
 
\begin{center}
\begin{figure}[t]
\includegraphics[width=6.5cm]{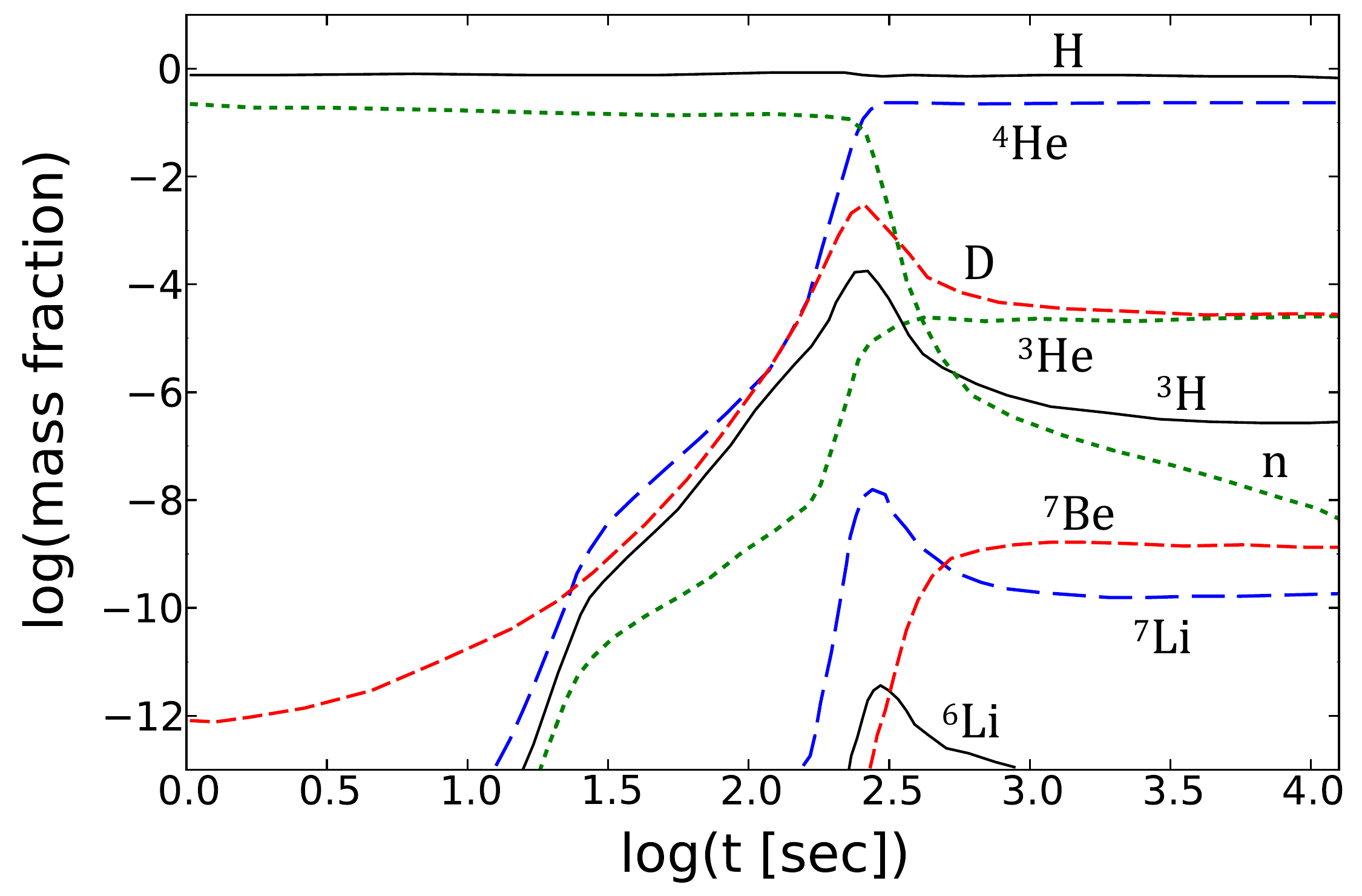}\ \ \ \ \ \ \ 
\includegraphics[width=6.cm]{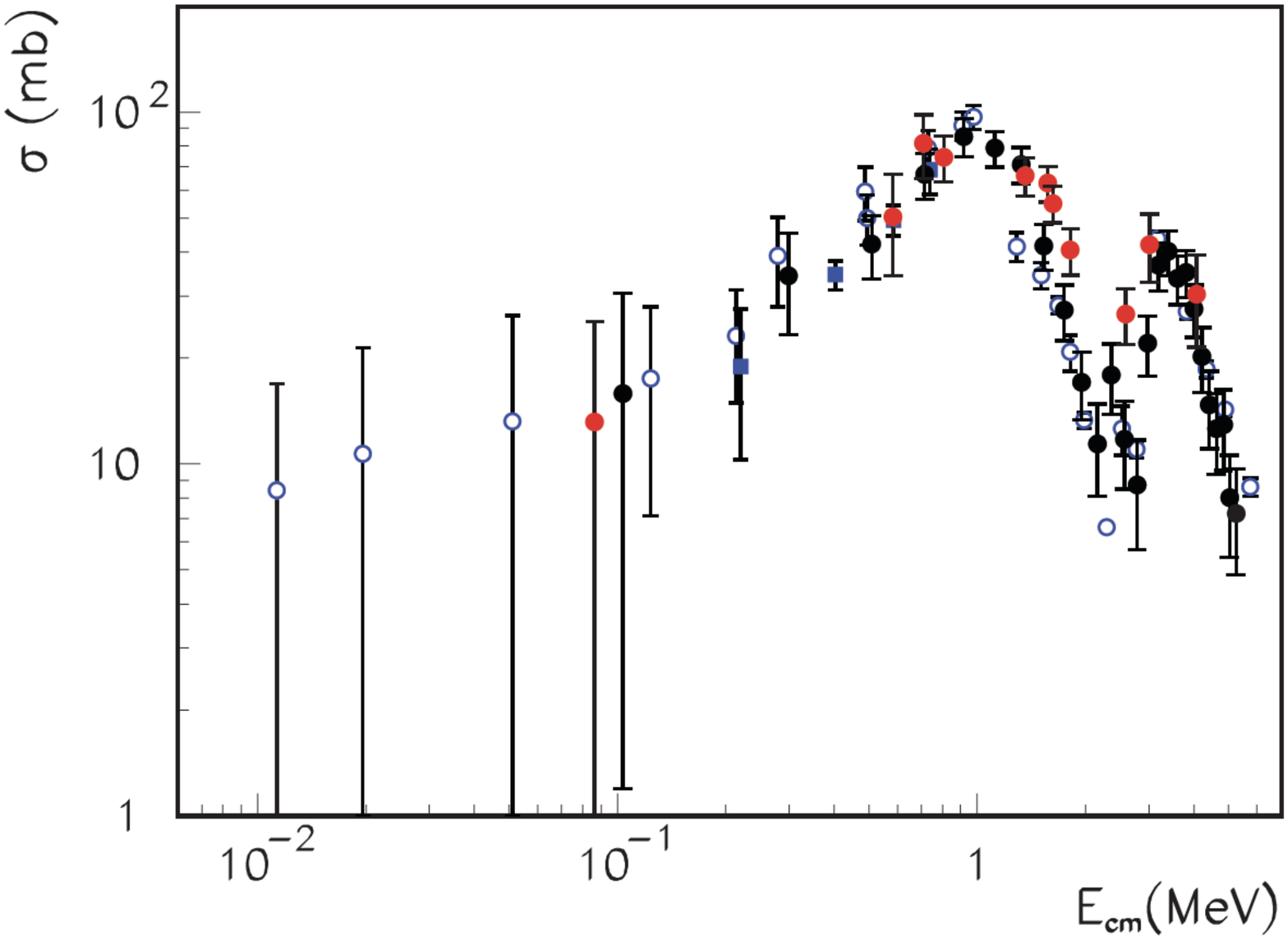}
\caption{{\it Left:} Calculated BBN abundances of H, D, $^3$H, $^3$He, $^4$He, $^6$Li, $^7$Li and $^7$Be as a function of time  \cite{BK16}.  {\it Right:} $^7$Be(n,$\alpha$) cross section deduced using the THM experimental data for the  $^7$Li(p,$\alpha)^4$He mirror reaction
(full red circles) and using $^3$He THM breakup data (full black circles) \cite{Lam17}. The data compiled by Hou \cite{Hou15} is shown as empty blue circles, and data from  Kawabata \cite{Kaw17} are shown as full blue squares. }\label{fig1}
\end{figure}
\end{center}

\section{Nuclear reaction cross sections}
\label{sec-2}
During the BBN, the most relevant nuclear reactions are listed in Table 1. This network of reactions resulted in the production of D, $^3$H, $^3$He, $^4$He, $^6$Li, $^7$Li and $^7$Be. Only very small traces of carbon, nitrogen and oxygen were produced at the $10^{-15}-10^{-25}$ abundance level. Therefore, there is no need to include reaction networks beyond those shown in Table 1, such as the famous CNO cycle, to tackle the lithium problem  \cite{Coc12}.  In Figure \ref{fig1} we show the calculated BBN abundance of H, D, $^3$H, $^3$He, $^4$He, $^6$Li, $^7$Li and $^7$Be as a function of time  \cite{BK16}. Our calculations were performed with an extended code  based on the Wagoner code \cite{Wag69} and similar to NUC123 \cite{Kaw92}. The dashed blue curve represents the $^4$He mass fraction, the red dashed curve represents the deuterium abundance, the green dashed curve represents the $^3$He abundance, the solid black curve is the $^3$H abundance, the red dashed curve is the $^7$Be abundance and blue dashed curve is the  $^7$Li abundance. Recently, new experimental measurements of reactions of relevance for the BBN have been reported based on the use of the Trojan Horse Method (THM) \cite{Piz14}.   

The reaction cross sections at the low astrophysical energies are enhanced due to the electrons in the plasma. The cross sections in the plasma are enhanced by a factor $  f(E) = {\sigma _{s}(E)/ \sigma _{b}(E)}$, where $\sigma _{s}$ is the screened and $\sigma _{b}$ the bare (non-screened) cross section. The  Debye-H\"uckel theory predicts a screened Coulomb potential of the form $V(r) = ({e^2 Z_i / r}) \exp\left(-{r/ R_D} \right)$, where the Debye radius is given by
$ R_D = (1/ \zeta)\left( {k T / 4 \pi e^2 n} \right)^{1/2}$, with $n$ being the ion number density and $ \zeta = \left[ \sum\limits_i X_i  (Z_i^2/A_i) +  \chi \sum\limits_i X_i (Z_i/A_i) \right]^{1/2}$, with $X_i$ the mass fraction of particle $i$ and the temperature $T_6$ in units of 10$^6$ K.  $\chi$ is a factor correcting for  electron degeneracy effects \cite{Sal54}. During the big bang, the electron number density decreased strongly with the temperature, being up to $10^4$ times larger than the number density in the core of the sun, $n_e^{sun}\sim 10^{26}$/cm$^3$.  However,  the baryon density was much smaller during the BBN epoch than at the core of the sun. The number of excess electrons during the BBN is nearly the same as those of protons. But most electrons were in balance with the number of positrons produced via  $\gamma \gamma \rightarrow e^+e^-$ processes.  In Ref. \cite{Wan11} the electron screening effects were included  in the BBN reaction network. It was found that the modification of the BBN abundances are negligible. Evidently, it cannot be responsible for the lithium abundance deficiency. Worth mentioning is that recently it has been shown that clustering effects in reactions involving light nuclei at astrophysically relevant energies might also play an important role and could perhaps explain some of the discrepancies found in the experimentally deduced values of electron screening enhancement and theoretical calculations \cite{Spit16}.

In Table \ref{tabbbn} the BBN calculations are compared with observations. The mass fraction for $^4$He, historically denoted by $Y_p$, is taken from Ref. \cite{YT10}, (b)  the deuterium abundance  ${\rm D/H} = (2.527 \pm 0.03) \times 10^{-5}$ \cite{RJ17,Coo18}, compatible with $100\Omega_b h^2 \ ({\rm BBN}) = 2.225 \pm 0.016$ inferred from the measurements of the cosmic microwave background \cite{Ade16}, (c)  the $^3$He abundance is taken  from Ref. \cite{BRB02},  and (d)  the lithium abundance  is taken from Ref. \cite{Sbo10}. The BBN model result for  the $^7$Li abundance shown in Table  \ref{tabbbn} is in evident discordance (roughly by a factor 3) with the observation. One possibility for this discrepancy could be that $^7$Be is further destroyed during the BBN. We recall that  $^7$Be decays  in $53.22 \pm 0.06$ days by electron capture  to ground state and the first excited state  (0.477 MeV) $^7$Li. Therefore, all $^7$Be  produced during the big bang will count towards the $^7$Li primordial abundance.  If $^7$Be is substantially destroyed by, e.g., (n,p) or (n,$\alpha$) reactions, it could possibly explain the observed lithium depletion. This possibility has been investigated in Refs. \cite{Brog12,Hou15,Bar16,Kaw17,Lam17}. In particular, Lamia et al. \cite{Lam17} have experimentally determined the  $^7$Be(n,$\alpha$) reaction cross section using the THM experimental data for the  $^7$Li(p,$\alpha)^4$He mirror reaction with corrections for Coulomb effects (see Figure \ref{fig1}). The new deduced data for $^7$Be(n,$\alpha$) using this technique lies within the Gamow window appropriate for BBN temperatures and the reaction rate using the new data is found to be lower by a factor $\approx 10$ relative to the one used by Wagoner \cite{Wag69}. The new reaction rate  yields a $^7$Li/H abundance ratio of $2.845 \times 10^{-11}$ and a $^7$Be/H abundance ratio of $4.156 \times 10^{-10}$, leading to a total  cosmological lithium abundance of $4.441 \times 10^{-10}$, and no appreciable change of the previously obtained BBN results for is verified. More recently, a theoretical investigation of the impact of the $^7$Be($\alpha,\gamma$) on $^7$Be destruction was performed \cite{Har18}. It was found that the $^7$Be abundance would be compromised only if an unexpected strong resonance exists very close to threshold in this reaction channel. All odds are that such a resonant state does not exist.

\begin{table}[htbp]
\vspace{0.0cm}
\centering
\caption{\label{tab:table2} BBN calculations using fits to recent experimental  data for BBN reactions compared with observations. The mass fraction for $^4$He, historically denoted by $Y_p$, is taken from Ref. \cite{YT10}, (b)  deuterium abundance  ${\rm D/H} = (2.527 \pm 0.03) \times 10^{-5}$ \cite{RJ17,Coo18}, compatible with $100\Omega_b h^2 \ ({\rm BBN}) = 2.225 \pm 0.016$ inferred from the measurements of the cosmic microwave background \cite{Ade16}, (c)  $^3$He abundance is taken  from Ref. \cite{BRB02},  and (d)  the lithium abundance  is taken from Ref. \cite{Sbo10}. 
}
\begin{tabular}{|c|c|c|c|c|c|c|c|}
\hline
\hline
Yields &   Calculation   & Observation\\  \hline
 
$Y_p$ &\text{0.2485{\footnotesize +0.001-0.002}}&$0.2565 \pm 0.006^{(a)}$ \\ \hline
D/H ($\times 10^{-5}$)&\text{2.692{\footnotesize +0.177-0.070}}&$2.527 \pm 0.03^{(b)}$ \\ \hline
${^3}$He/H ($\times 10^{-6}$) &\text{9.441{\footnotesize +0.511-0.466}}&$\geq 11.\pm 2.^{(c)}$\\ \hline
${^7}$Li/H ($\times 10^{-10}$)&\text{4.283 {\footnotesize +0.335-0.292}}&$\text{1.58{\footnotesize +0.35 -0.28}}^{(d)} $ \\ \hline
\hline
\end{tabular} 
\vspace{0.0cm}
\label{tabbbn}
\end{table}

BBN predicts an isotopic ratio of $^6{\rm Li}/^7{\rm Li} \sim 10^{-5}$, whereas  observation yields $^6{\rm Li}/^7{\rm Li} \sim 2 \times 10^{-2}$ \cite{Asp06}. In Ref. \cite{MSB16} a re-analysis of the reaction $^4$He$({\rm d},\gamma)^6$Li was performed, including new predictions for the gamma-ray angular distribution. This was done using a two-body potential model to calculate the S-factor for this reaction at the BBN energies \cite{radcap,MSB16}. The potential parameters were chosen to reproduce experimental phase shifts and recently measured ANCs.  A nice agreement was found with the experimental data of the LUNA collaboration \cite{LUNA}.  This work reinforces BBN predictions for the lithium isotopic ratio and yields a new value of $^6{\rm Li}/^7{\rm Li} = (1.5 \pm 0.3) \times 10^{-5}$. The second lithium puzzle seems to be alive although it is not impossible that  lithium abundances might change appreciably due to astration.

We conclude this section by stating that it does not seem possible that both lithium puzzles can be solved by accurate measurements  of nuclear reaction cross sections, combined with progresses in the theories for nuclear astrophysical reactions. There has been a considerable number of recent theoretical efforts to elucidate the lithium puzzle using a plethora of different ideas based on the premise that physics as we know today might have been different 13.8 billions years ago. New particles, new interactions, changes in fundamental constants, non-standard BBN models, and various intriguing ideas have been used and published elsewhere.       
 
\section{Dark matter}

Most of the matter in the universe consists of an obscure kind of Dark Matter (DM)  which interacts very weakly with the visible matter. In fact, we only know that it interacts gravitationally and large scale experimental searches are underway to identify if DM interacts with visible matter by other means
 \cite{Feng2010,bertone05,bertone10}.  The existence of DM is based on astronomical observations of galaxy clusters dynamics and on the anisotropies of the Cosmic Microwave Background (CMB). Perhaps  Weakly Interacting Massive Particles (WIMPs), supersymmetric particles, sterile neutrinos, or any other hitherto undiscovered particles are responsible for its composition. It has also been hypothesized that  DM is a mirror sector of particles such as dark photons, dark electrons, etc., which interact in nearly the same way as Standard Model (SM) particles, but only within their own sector. They interact very weakly across sectors, i.e. between the DM sector and the visible sector \cite{Yang,Kobzarev,Pavsic,Foot,Akhmedov,Oli11,Oli16}. Besides, the particle copies in the dark sector do not need to have the same masses and couplings as in the visible sector, opening a huge number of possible scenarios for DM. 

Astronomical observations yield the ratio of density parameters $\Omega_{DM} / \Omega_{visible} =4.94 \pm 0.66$. Therefore, DM is 5 times more frequent than visible matter.  In Ref. \cite{Oli11,Ber13,Oli16} this feature was used to explore the possible existence of 5 dark sectors instead of the single ubiquitous dark sector.  An Weakly Interacting Massive Gauge Boson (WIMG) was also proposed to couple all dark sectors and ordinary matter. The massive, $E\sim 10$ TeV,  WIMG  does not modify the properties of the SM and gravity. It has to be consistent with BBN predictions and CMB observations, except maybe with the lithium abundance.  Much below the electroweak scale energy, we can assume particles to be massless and group them in matter/charge fields with a similar structure for DM. The WIMG mass is generated by a   real scalar field, with the condition that the WIMG  has a short-range interaction. In this formalism, the number of dark sectors plays an important role which has been overseen in other BBN models. The new  degrees of freedom of particles in the dark sector modify the early universe expansion rate  \cite{Berezhiani1996} and the elemental abundance predictions. Additional dark sectors increase the effective number  degrees of freedom and their implications for BBN  \cite{Oli11,Ber13,Oli16}.   

The basic idea of having additional dark sectors is that the radiation density in the BBN epoch have densities and entropies given by
$  \rho(T)=({\pi^{2}}/{30}) \, g_*(T) \,T^4 $ and $ s(T)=({2\pi^{2}}/{45}) \, g_s(T)\, T^3$ with
\begin{equation}
g_*(T)=\sum_B g_B \left(\frac{T_{B}}{T}\right)^4 + \frac{7}{8} \sum_F g_F \left(\frac{T_{F}}{T}\right)^4, \ \ \ \ 
{\rm and} \ \ \ \
g_s(T)=\sum_B g_B \left(\frac{T_{B}}{T}\right)^3 + \frac{7}{8} \sum_F g_F \left(\frac{T_{F}}{T}\right)^3, \label{dof}
\end{equation}
where $g_*$ and $g_s$ are the number of degrees of freedom, with $g_{B(F)}$ being the fractions contributed by  bosons (fermions) at temperatures $T_{B(F)}$. In this notation, $T$ is the temperature of the radiation thermal bath.

\begin{figure}[ptb]
\begin{center}
\includegraphics[width=4.5cm]{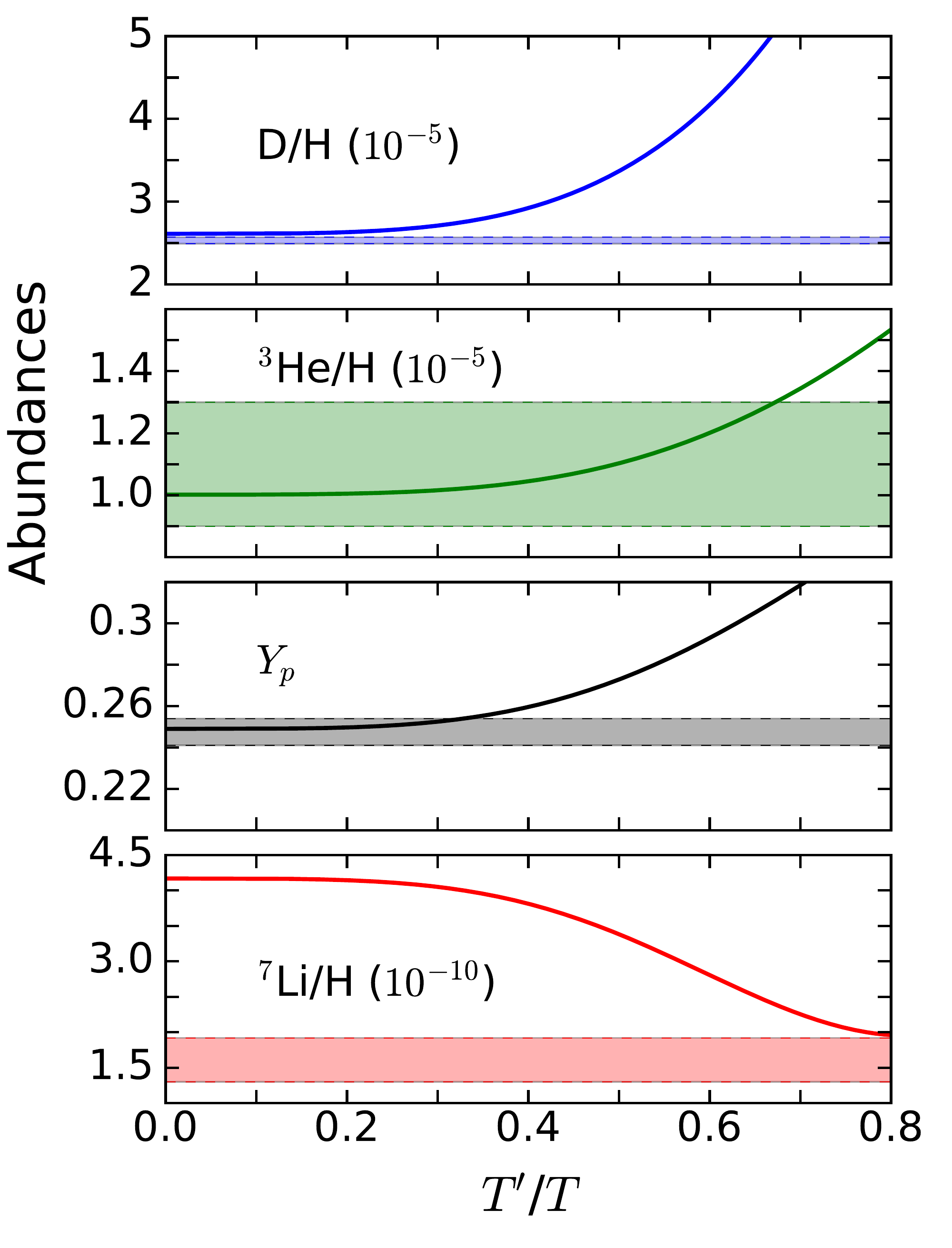}\ \ \ \ \ \ \ 
\includegraphics[width=6.cm]{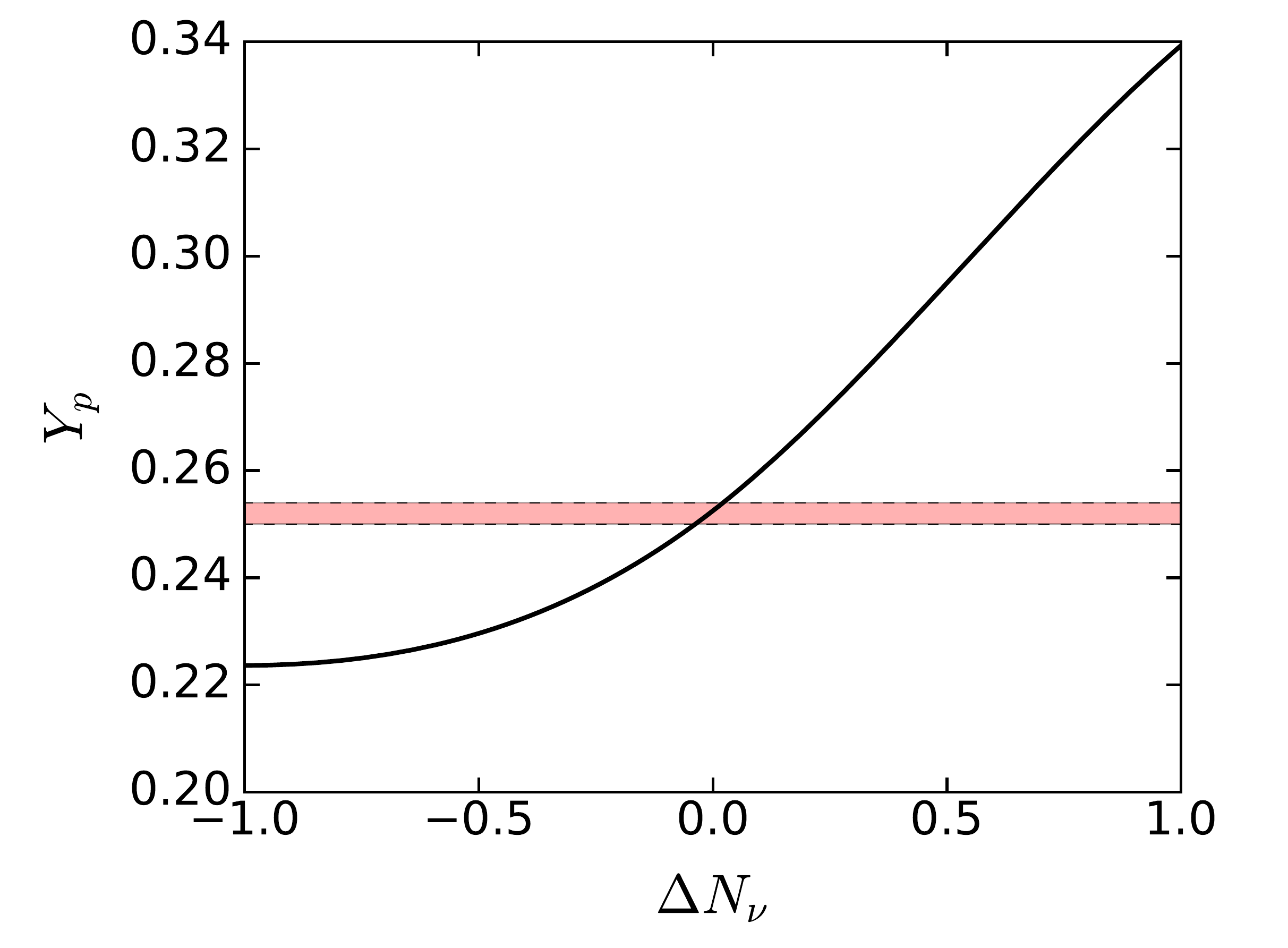}
\end{center}
\caption{{\it Left:} Calculated relative abundances of D, $^3$He, $^4$He (mass fraction, $Y_p$) and $^7$Li as a function of $T'/T$, with $N_{DM}=5$ compared to observations. The bands represent uncertainties in the observations \cite{Ber17}. {\it Right:} Predictions for the primordial $^4$He mass fraction as a function of extra  neutrino families, with $T'= 0.3T_{BBN}$ and  $N_{DM}=5$. The horizontal band represents the observed mass fraction \cite{Ber17}.}
\label{fig2}
\end{figure}

For simplicity, we assume only two temperatures: $T$ in the ordinary matter sector and $T'$ in the dark sectors. By analogy, the
energy $\rho'(T')$ and entropy $s'(T')$ densities in the dark sectors are also obtained with Eqs. (\ref{dof}) with $g_*(T) \rightarrow g'_*(T')$, $g_s(T) \rightarrow g'_s(T')$, and  $T\rightarrow T'$. An independent variable $x=(s'/s)^{1/3}\sim T'/T$ emerges if one assumes conservation of entropy in all sectors. If each dark sector has the same matter content as in the visible sector,  then $g_s(T_0) = g_s^\prime(T'_0)$, leading to $x=T'/T$. The Friedman equation is $ H(t)=\sqrt{\left(8\pi/3 c^2\right) \, G_{N} \, \bar{\rho}}$, where $\bar{\rho}$ is the total energy density. Including the number of dark sectors, $N_{DM}$ it becomes  $\bar{\rho} = \rho\, + \, N_{DM} \, \rho'$. Therefore, one has
$H(t)=1.66 \, \sqrt{\bar{g}_{*}(T)} {T^2}/{M_{Pl}}$, with $\bar{g}_{*}(T) = g_{*} (T) \left( 1+ N_{DM} \, a \, x^4 \right)$,
where $M_{Pl}$ is the  Planck mass and $a = \left(g'_{*}/g_{*}\right) \left(g_{s}/g'_{s}\right)^{4/3}\sim 1$, for a not too small $T'/T$  \cite{Berezhiani1996}. At  about 1 MeV,  standard BBN assumes $g_{*}(T=1 \ {\rm MeV}) = 10.75$, but with the  additional dark particles it becomes  $\bar{g}_{*} = g_{*} \left(1 + N_{DM} \, x^4 \right)$. We can study the bounds for $N_{DM}$ and $x$, or $T'/T$, by comparing BBN calculations and the relative abundances of the light element isotopes (D, $^{3}$He, $^{4}$He, and $^{7}$Li).  This is shown in Figure \ref{fig2} as a function of $T'/T$ with a fixed number of dark sectors, $N_{DM}=5$. The shaded bands include the uncertainty in the observed values. In this case, we notice that observations of primordial elements of D, $^{3}$He, and $^{4}$He  are compatible with $T'/T\sim 0.2-0.3$. 

The $^7$Li problem remains because if $T'/T\sim 1$ then $^7$Li comes out right, but the other abundances will be completely off the observations.  $\bar{g}_{*}(T)$ is much more sensitive to $T'$ than it is to $N_{DM}$. Using $T'=0.3T$  for cold dark sectors, we obtain a large range of values for  $N_{DM}=1-50$ compatible with the D, $^3$He and $^4$He  abundances \cite{Ber17}.  Figure \ref{fig2} also shows the predictions for the primordial $^4$He mass fraction as a function of extra  neutrino families, $\Delta N_\nu$, with $T'= 0.3T_{BBN}$ and  $N_{DM}=5$. The horizontal band represents the observed mass fraction \cite{Ber17}. The model is thus compatible with the number of neutrino families $N_\nu =3$. We thus conclude that there is no incompatibility with the observed primordial abundances and a universe composed with more than one sector of dark matter, e.g. $N_{DM} =5$ and temperatures of the dark sectors of the order of $T'\sim 0.2-0.3T$.

\section{Non-extensive statistics}

The Maxwell-Boltzmann (MB) distribution  is widely known to reproduce extremely well the distribution of velocities of particles in a thermal bath.   The MB distribution is a result of the Boltzmann-Gibbs statistics, based on the assumptions that (a) the time between collisions among particles is much larger than their interaction time, (b) the interaction is short-ranged, (c) no correlation exists between the particle velocities, and (d) the collision energy is conserved without transfer to internal degrees of freedom. These very constraining assumptions are not expected to be always valid in thermodynamical equilibrium.  In fact, alternatives to the Boltzmann-Gibbs (BG) statistics are known to exist \cite{Ren60,Ts88,GT04}. In Ref. \cite{BFH13}, one of these non-extensive statistics, namely, the Tsallis statistics \cite{Ts88,GT04} has been used to describe the relative velocities of particles during the  BBN. The effect on the lithium abundance was again the motivation for this work. The Tsallis statistics was used \cite{Ts88,GT04}, because it represents a family of entropies depending on a parameter $q$, which measures the departure from Boltzmann statistics. The Boltzmann statistics is recovered when $q = 1$.

\begin{figure}[ptb]
\begin{center}
\includegraphics[width=7.cm]{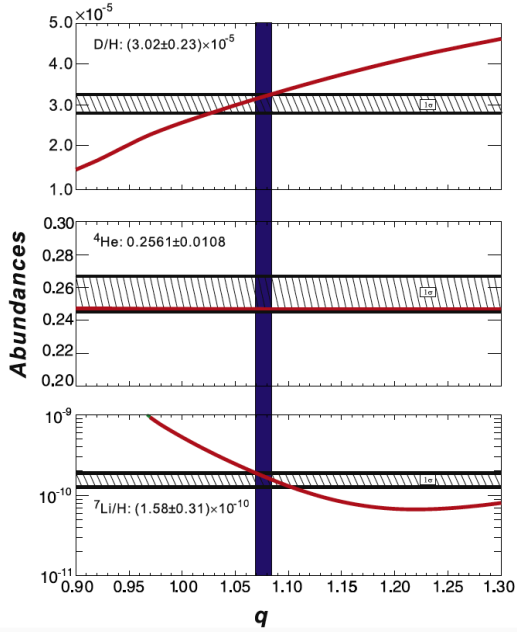}
\end{center}
\caption{{\it Left:} Predicted abundances D, $^4$He and $^7$Li (red curves) as a function of the Tsallis parameter $q$  \cite{Hou17}. The observed primordial abundances including 1$\sigma$ unvertainties are indicated by hatched horizontal bands \cite{Ave10,Sbo10,Oliv12}. The vertical (blue) band refers to the parameter $q$ within the interval $1.069<q<1.082$.}
\label{fig3}
\end{figure}

In all previous applications of non-extensive statistics, it has been found that the non-extensive parameter $q$ does not depart appreciably for the Boltzmann value $q=1$.  Non-extensive Maxwellian velocity distributions have previously been applied to study stellar nuclear burning, e.g. in Refs.  \cite{MQ05,HK08,Deg98,Cor99}.  In Ref. \cite{BFH13} the Tsallis statistics was used to deduce reaction rates during the  BBN  and predictions were made for the  ${^4}$He, D, ${^3}$He, and ${^7}$Li abundances which are based on the reaction rates for p(n,$\gamma$)d, d(p,$\gamma){^3}$He, d(d,n)${^3}$He, d(d,p)t, ${^3}$He(n,p)t, t(d,n)${^4}$He, ${^3}$He(d,p)${^4}$He, ${^3}$He$(\alpha,\gamma){^7}$Be, t$(\alpha,\gamma){^7}$Li, ${^7}$Be(n,p)${^7}$Li and ${^7}$Li(p,$\alpha){^4}$He and their available experimental data \cite{BFH13}.  The conclusion from Ref.  \cite{BFH13} is that if either $q>1$ of $q<1$, the abundances of all elements are affected but that of $^7$Li always increases. Therefore, it was inferred that the lithium problem always seems to get worse with the use of the Tsallis statistics.

However, there was a small, but relevant point neglected in the calculations of Ref. \cite{BFH13} was the proper inclusion of the reaction Q-values in the reaction rates obtained with the Tsallis statistics. This was fixed in Ref.  \cite{Hou17} and shown that, when the Q-values for the reverse reactions are properly accounted for, a beautiful result emerges for a relatively small departure of the parameter $q$ from the unity. The abundances of  H, D, $^3$H, $^3$He, and $^4$He, do not change, but that of $^7$Li does change appreciably, and in the correct direction to solve the $^7$Li puzzle.  in fact, an excellent agreement was found between the calculated and the primordial abundances observed for D, $^4$He, and $^7$Li for $1.069<q<1.082$, indicating that a possible solution to the cosmological lithium problem might arise from a fine tuning of the physics involved. This is shown in Figure  \ref{fig3} with  the predicted abundances D, $^4$He and $^7$Li (red curves) as a function of the Tsallis parameter $q$. The observed primordial abundances including 1$\sigma$ uncertainties are indicated by hatched horizontal bands \cite{Ave10,Sbo10,Oliv12}. The vertical (blue) band refers to the parameter $q$ within the interval $1.069<q<1.082$.

The work published in Ref. \cite{Hou17} was cited as a research highlight by the American Astronomical Society \cite{AAS17}. It attests the relevance of the lithium puzzle and the anxiety that its solution entails for the astronomical community. The puzzle has been around the literature for a few decades already. The exercise played in Ref. \cite{Hou17} shows that a solution might be the outcome of a fine tuning of the physics during the BBN. The Tsallis statistics might be one possible departure from the standard physics during the big bang epoch.   The question remains on the physical meaning for the value of $q\ne 1$, and its relation to other physical processes.

\medskip

{\bf Acknowledgements}

This work was supported in part by the U.S. DOE grant DE- FG02-08ER41533 and the U.S. National Science Foundation Grant No. 1415656.

\end{document}